%% file: main.tex
\documentclass[sigconf]{acmart}
\setcopyright{none}
\renewcommand\footnotetextcopyrightpermission[1]{} 

\usepackage[english]{babel}
\usepackage[utf8]{inputenc}
\usepackage{amsthm}
\usepackage{amsmath}
\usepackage{bbm}
\usepackage{bm}
\usepackage{mathtools}
\usepackage{graphicx}
\usepackage{caption}
\usepackage{subcaption}
\usepackage{comment}
\usepackage{xcolor}
\usepackage{enumitem}
\usepackage{hyperref}

\DeclareMathOperator*{\argmax}{arg\,max}
\DeclareMathOperator*{\argmin}{arg\,min}
\newcommand{\abs}[1]{\lvert{#1}\rvert}

\newcommand{\mc}[1]{\mathcal{#1}}
\newcommand{\mb}[1]{\mathbf{#1}}
\newcommand{\nin}{n \in \mc{N}}
\newcommand{\kin}{k \in \mc{K}}
\newcommand{\jin}{j \in \mc{J}_n}
\newcommand{\iin}{i \in \mc{I}_n}

\newcommand{\kjin}{(k,j) \in \mc{B}_{n,i}}
\newcommand{\kjout}{(k,j) \not\in \mc{B}_{n,i}}
\newcommand{\ain}{a \in \mc{N}}
\newcommand{\bin}{b \in \mc{N}}
\newcommand{\abin}{(a,b) \in \mc{L}}
\newcommand{\about}{(a,b) \not\in \mc{L}^k}
\newcommand{\drift}{\Delta(\mathbf{V}(t))}
\newcommand{\pen}{\mathbbm{E}[p(t)|\mathbf{V}(t)]}
\newcommand{\penhat}{\mathbbm{E}[\hat{p}(\svec)|\mb{V}(t)]}
\newcommand{\pentilde}{\mathbbm{E}[\hat{p}(\svectilde)|\mathbf{V}(t)]}
\newcommand{\penstar}{\mathbbm{E}[\hat{p}(\svecstar)|\mathbf{V}(t)]}
\newcommand{\flowan}{\sum\limits_{\ain}\mu^k_{an}(t)}
\newcommand{\flownb}{\sum\limits_{\bin}\mu^k_{nb}(t)}
\newcommand{\flowantilde}{\sum\limits_{\ain}\tilde{\mu}^k_{an}(t)}
\newcommand{\flownbtilde}{\sum\limits_{\bin}\tilde{\mu}^k_{nb}(t)}
\newcommand{\betasum}{\sum\limits ^{\sigma}_{i=1} \beta_{n,i}}

\newcommand{\betasumind}{\sum\limits ^{\sigma}_{i=1} \beta_{n,i}\mathbf{1}_{[\kjin]}}

\newcommand{\rosumtier}{\sum\limits_{\jin} r_{n_j}s^k_{n_j}(t)}
\newcommand{\rosumtiertilde}{\sum\limits_{\jin} r_{n_j}\tilde{s}^k_{n_j}(t)}

\newcommand{\minpen}{\Psi(\boldsymbol{\lambda})}
\newcommand{\svec}{\textbf{\textit{s}}(t)}
\newcommand{\svectilde}{\tilde{\textbf{\textit{s}}}(t)}
\newcommand{\svecstar}{\textbf{\textit{s}}^*(t)}

\newtheorem{algorithm}{Algorithm}
\newtheorem{theorem}{Theorem}
\newtheorem{lemma}{Lemma}

\makeatother

\begin{document}
\title{Cost-aware Joint Caching and Forwarding in Networks with Heterogeneous Cache Resources}

\author{Faruk Volkan Mutlu}
\email{fvmutlu@ece.neu.edu}
\affiliation{%
  \institution{Northeastern University}
  \city{Boston}
  \state{MA}
  \country{USA}
}

\author{Edmund Yeh}
\email{eyeh@ece.neu.edu}
\affiliation{%
  \institution{Northeastern University}
  \city{Boston}
  \state{MA}
  \country{USA}
}

\begin{abstract}
\input{Body/abstract}
\end{abstract}

\maketitle
\pagestyle{plain} 

\section{Introduction}
\label{sec:s1_intro}
\input{Body/S1_intro}

\section{Related Work}
\label{sec:s2_relatedwork}
\input{Body/S2_related}

\section{System Model}
\label{sec:s3_model}
\input{Body/S3_model}

\section{Control Plane}
\label{sec:s4_control}
\input{Body/S4_control}

\section{Data Plane}
\label{sec:s5_data}

\input{Body/S5_data}

\section{Experimental Results}
\label{sec:s6_results}
\input{Body/S6_numerical}

\section{Conclusion and Future Work}
\label{sec:s8_conc}
\input{Body/S8_conclusion}

\bibliographystyle{ACM-Reference-Format}
\bibliography{references} 

\appendix
\section{Proof of Theorem 1}
\label{app:stability}
\input{Appendices/A1_stabilityregion}

\section{Proof of Theorem 2}
\label{app:dpp}
\input{Appendices/A2_minimaldpp}

\end{document}

%% file: Body/abstract.tex
Caching is crucial for enabling high-throughput networks for data intensive applications.  Traditional caching technology relies on DRAM, as it can transfer data at a high rate. However, DRAM capacity is subject to contention by most system components and thus is very limited, implying that DRAM-only caches cannot scale to meet growing demand. Fortunately, persistent memory and flash storage technologies are rapidly evolving and can be utilized alongside DRAM to increase cache capacities. To do so without compromising network performance requires caching techniques adapted to the characteristics of these technologies.
In this paper, we model the cache as a collection of storage blocks with different rate parameters and utilization costs. We introduce an optimization technique based on the drift-plus-penalty method and apply it in a framework which enables joint caching and forwarding.  We show that it achieves an optimal trade-off between throughput and cache utilization costs in a virtual control plane.  We then develop a corresponding practical policy in the data plane. Finally, through simulations in several settings, we demonstrate the superior performance of our proposed approach with respect to total user delay and cache utilization costs.

%% file: Body/S1_intro.tex
One of the core components enabling high-throughput data delivery networks is in-network caching. The main parameters of a cache that impact network performance are its transfer rate and capacity. The traditional cache device is DRAM, which despite its small capacity, can match network link rates. However, as the volume of data grows, the small capacity of DRAM becomes an obstacle in scaling network performance via caching.
Fortunately, persistent storage technologies are advancing rapidly and they may soon have transfer rates high enough to justify their use as additional tiers of cache alongside DRAM to enable larger capacities.
System designs that consider this {\it vertical} scaling of caches already exist in the literature, and given the expected performance improvements in upcoming evolutions of the PCIe and NVMe standards, as well as technologies like SPDK \cite{yang2017spdk}, wider implementation of routers with large multi-tiered caches is imminent.

Nonetheless, a transfer rate gap between memory and storage will continue to exist for the foreseeable future, even if it narrows over time. Furthermore, while storage technologies today offer large capacities at low upfront costs, utilizing them as short-term caches introduce significant operational costs, as they consume significant amounts of power and their reliability diminishes with use. Therefore, the benefits and costs of using these devices must be carefully balanced in order to build sustainable high performance networks. The first step in this balancing act is the cache admission and replacement policy. However, policies in wide adoption today tend to assume a singular cache capable of operating at the rate of the forwarding path and emphasize the characteristics of data objects rather than caches, making them unsuitable for the task.

In contrast, a {\it multi-tiered caching policy} maintains the goal of caching a small set of data objects with high demand in the fastest tier of cache, but also aims to collect a larger set of objects with less demand in slower tiers with more capacity. The objects in the slower tiers are requested less often, but there are a larger number of them. Therefore, such a policy aims to balance the frequency of cache hits served by each tier based on their transfer rates. Furthermore, it makes cache replacement decisions carefully to mitigate the number of low benefit replacements, where a data object is replaced with another one of only slightly more demand. Such replacements are costly, happen more frequently in slower tiers, and bring no significant performance gains in the long term. In this paper, we propose a multi-tiered caching policy that can intelligently utilize a variety of memory and storage elements with different characteristics. The goal of this policy is to find an optimal trade-off between the performance gained via caching and the costs incurred by utilizing caches.  The proposed policy is built on the VIP framework \cite{VIPoriginal}, which provides powerful design and analysis tools that fit our goals.
The contributions of this paper can be summarized as follows:
\begin{itemize}[leftmargin=5mm]
    \item An object-level caching model that considers multiple tiers of cache with different cost and performance parameters, and an extension of the VIP framework to this caching model,
    \item A distributed joint caching and forwarding algorithm built in the virtual control plane of this extended framework which optimally balances queue growth bound and cache utilization costs, and an exact solution to the optimal cache placement problem posed by this algorithm,
    \item Practical caching and forwarding policies in the data plane driven by the virtual control plane algorithm and a thorough experimental evaluation of these policies through simulations.
\end{itemize}

%% file: Body/S2_related.tex
The challenge of optimal data placement in multi-layered storage architectures have long been a focus of research in systems like data centers or cloud storage platforms. In these contexts, vast amounts of data need to be kept in storage for long periods of time. To provide low latency access to popular portions of this data, techniques that use SSDs as caches for HDDs are widely adopted \cite{niu2018hybrid}. Novel techniques aimed at modern enterprise storage systems where all storage elements are SSDs (all-flash) have also been developed \cite{yang2017autotiering}. Most recently, reinforcement learning was applied to the data placement problem in hybrid storage systems to provide an adaptive and continuously improving solution \cite{singh2022sibyl}.

The optimal data placement challenge for the in-network caching context may appear similar. However, the crucial difference is that in-network caching places cache devices directly on the forwarding path of requests and data. Therefore, bottlenecks in the caching system lead to compounding network-wide performance issues. As such, in-network caching has much stricter demands from storage systems and lower tolerance for overheads.
Still, the feasibility of devices beyond DRAM as caches in a data delivery network context have been explored, primarily for the information-centric networking (ICN) paradigm.
While SSDs were initially seen as either too costly or not performant enough for ICN \cite{perino2011reality}, as storage technologies improved, system designs that make use of them for ICN caches appeared.
In \cite{rossini2014multi}, a two-layer cache with a large SSD layer masked behind a fast DRAM layer was designed, which was later used in a follow-up work to build a high-speed router prototype which achieved more than 10Gbps throughput \cite{mansilha2015hierarchical}. Another design was proposed in \cite{so2014toward}, where block device I/O was separated from forwarding paths by sending packets to a dedicated caching process, allowing forwarding to continue operating at line-rate.

However, while such designs address the integration of persistent storage into caching systems, they are not backed by policies designed for multi-tiered caches and the constraints of in-network caching. The lack of such policies is one of the reasons why SSD-based caches have yet to see wider adoption in modern high-performance ICN prototypes.
In \cite{shi2020ndn}, where a high-throughput software-defined forwarder for Named Data Networking (NDN) was introduced, caching in persistent memory or NVMe disk storage was stated as a future direction and the necessity for novel caching algorithms accounting for varying device characteristics in a multi-tiered cache was emphasized. Most recently, an experimental study for a NDN-based data delivery system, intended for science experiments with data volumes in the exabyte range, featured nodes with DRAM-only caches of 20~GB capacity \cite{wu2022ndise}.

While neither traditional policies nor emerging ones from the literature readily propose extensions to multi-tiered caches, the VIP framework described in \cite{VIPoriginal} stands out as a primary candidate that can support such an extension. Although the algorithm proposed in \cite{VIPoriginal} targets single-tier caches, the framework models the transfer rate of that cache. In Section~\ref{sec:s4_control}, we describe how we use this fact to extend the framework to multiple tiers.

%% file: Body/S3_model.tex
Let the unit of content in the network be a \textit{data object} (or simply {\it object} for short) and assume that each object has the same size\footnote{While this assumption simplifies our model and analysis, the approach presented in this paper can be generalized to a model with data objects of unequal size. Section~\ref{subsubsec:solution} contains further discussion on this issue.}. Denote the set of objects in the network as $\mc{K}$. Let $\mc{G} = (\mc{N},\mc{L})$ be a directed graph where $\mc{N}$ and $\mc{L}$ denote the sets of nodes and links in the network respectively. Assume that $(b,a) \in \mc{L}$, if $\abin$ and let $C_{ab}$ be the transmission capacity of link $\abin$ in objects/second. For each object $\kin$, assume that there is a unique and fixed node $\mc{S}(k) \in \mc{N}$ which serves as the content source of $k$. Requests for data objects can enter the network at any node and are transmitted via request packets of negligible size. A request for object $k$ is forwarded through the network until it reaches either $\mc{S}(k)$ or a node $n$ that caches $k$, at which point the data object is produced at $n$ and delivered to the requester, following the path of the request in reverse.

Assume that any node in the network can have one or more cache(s), each referred to as a {\it cache tier}. Denote the set of cache tiers at node $\nin$ as $\mc{J}_n$ and let tiers be indexed as $j=1,2,...$, i.e. $\jin = \{1,2...\}$. Let each tier have capacity $L_{n_j}$ in objects. Assume that an object $k$ can be cached in at most one tier at a node $n$, and cannot be cached in any of the tiers at node $\mc{S}(k)$. We refer to this as the {\it cache exclusivity} assumption.
Migrations of objects from one cache tier to another occur when a replacement takes place in one tier and the evicted object is moved into another tier, instead of being evicted entirely from the node. To facilitate such migrations, we assume that each node has a {\it migration buffer} capable of storing one object at a time, where a migrating object is kept temporarily. While this is only required when the destination tier is full, for simplicity we assume that a migrating object is always moved to this buffer first.
Admitting a new object into tier $j$ at node $n$ incurs the cost $c^a_{n_j}$. This can represent the time or energy spent on this write operation, its endurance impact, or a combination of these and any other potential costs. Evicting an object from tier $j$ and placing it in the migration buffer incurs the cost $c^e_{n_j}$, which includes both the cost of reading from $j$ and that of writing to the buffer.

%% file: Body/S4_control.tex
As stated in Section~\ref{sec:s1_intro}, we build our multi-tiered caching policy on the VIP framework \cite{VIPoriginal}. In this framework, user demand for objects is measured using \textit{virtual interest packets} (VIPs), which are tracked in a {\it virtual control plane} (or simply {\it virtual plane}). The virtual plane operates separately from the request forwarding and data delivery processes described in Section~\ref{sec:s3_model}, which take place in what we refer to as the {\it data plane}.
This approach facilitates the development of joint caching and forwarding algorithms in the virtual plane. VIPs are used as the common metric that measures the value of data objects, which allow such algorithms to drive requests toward caching nodes while avoiding congestion in the network.
Additionally, this separation reduces the complexity of design and analysis for such algorithms, compared to building them directly in the data plane.

In view of our design goal described in Section~\ref{sec:s1_intro}, the VIP framework provides two additional advantages. First, it embeds the most critical parameter of a cache, the rate at which it can push objects into the network, directly into its analysis of network dynamics in the virtual plane. Secondly, as this analysis is based on the {\it Lyapunov drift} technique, it allows us to formulate a performance-cost trade-off using the {\it drift-plus-penalty} approach \cite{neelybook}.

In this section, we describe the dynamics of VIPs, introduce our virtual plane joint caching and forwarding algorithm and show that this algorithm achieves an optimal trade-off between VIP backlogs and cache utilization costs.

\subsection{VIP Dynamics}
\label{subsec:vipdynamics}

Time in the virtual plane is represented in slots of length 1 indexed as $t =  1,2, ...$ where time slot $t$ refers to the time interval $[t,t+1)$.
Each node $\nin$ keeps a separate counter $V^k_n(t)$ for each object $\kin$, which are all set to 0 at the beginning of time slot $t = 1$, with $V^k_n(t) = 0$ for all $t \geq 1$ if $n = \mc{S}(k)$. These are called {\it VIP counts}.
When a request for object $k$ enters the network at node $n$ in the data plane, the corresponding counter $V^k_n(t)$ is incremented. The number of requests for object $k$ that enter the network at node $n$ during slot $t$ is denoted as $A^k_n(t)$ and the external {\it arrival rate} of VIPs at node $n$ for object $k$ is $\lambda^k_n \triangleq \lim_{t \rightarrow \infty} \frac{1}{t}\sum^t_{\tau = 1} A^k_n(\tau)$.
At each time slot, nodes communicate their VIP counts and forward VIPs to their neighbors according to decisions made by a virtual plane algorithm. The allocated transmission rate of VIPs for object $k$ over link $(a,b)$ during slot $t$ is denoted as $\mu^k_{ab}(t)$. VIPs for an object $k$, after being forwarded through the virtual plane, are removed at either $\mc{S}(k)$ or nodes that cache $k$. Note that, in the data plane, the communication of VIPs can be handled with a single message of negligible size at each slot.
We define the {\it cache state} for object $k$ at node $n$ in tier $j$ during slot $t$ as $s^k_{n_j}(t) \in \{0,1\}$, where $s^k_{n_j}(t) = 1$ if the object is cached and $s^k_{n_j}(t) = 0$ otherwise, with $s^k_{n_j}(1) = 0, \; \forall \nin,\kin$.
We assume that in the virtual plane, at each time $t$, a node can immediately gain access to any data object in the network and cache it locally.
A tier $j$ at node $n$ has {\it readout rate} $r_{n_j}$, meaning it can produce $r_{n_j}$ copies of each object it caches per time slot independent of other objects and tiers. We assume the cache tiers in the set are numbered in descending order of their readout rates, i.e. $r_{n_1} \geq r_{n_2} \geq \cdots$.

\subsection{Joint Caching and Forwarding Algorithm}
\label{subsec:vipalgo}

We now describe our distributed joint caching and forwarding algorithm for the virtual plane in terms of the definitions above.
\begin{algorithm}[Caching and Forwarding in the Virtual Plane]
At the beginning of each time slot $t$, at each node $n$, observe VIP counts $(V^k_n(t))_{\kin, \nin}$, then perform forwarding and caching in the virtual plane as follows. \\
{\bf Caching:} At each node $\nin$, choose $s^k_{n_j}(t)$ for each data object $\kin$ and cache tier $\jin$ to
\begin{align}
    \text{maximize}
        & \quad \sum\limits_{\kin} \sum\limits_{\jin} b^k_{n_j}(t) s^k_{n_j}(t) \label{eq:gap_obj} \\
    \text{subject to}
        & \quad \sum\limits_{\kin} s^k_{n_j}(t) \leq L_{n_j}, \; \jin \label{eq:gap_const1} \\
        & \quad \sum\limits_{\jin} s^k_{n_j}(t) \leq 1, \; \kin \label{eq:gap_const2} \\
        & \quad s^k_{n_j}(t) \in \{0, 1\}, \; \kin, \; \jin \label{eq:gap_const3}
\end{align}
where,
\begin{equation}
    b^k_{n_j}(t) \triangleq \left\{ \begin{array}{ll}
        r_{n_j} V^k_n(t) - \omega c^a_{n_j}, & \text{if} \; s^k_{n_j}(t-1) = 0 \\
        r_{n_j} V^k_n(t) + \omega c^e_{n_j}, & \text{if} \; s^k_{n_j}(t-1) = 1
    \end{array} \right.
\label{eq:gap_benefits}
\end{equation}
and $\omega \geq 0$ is the importance weight of the penalty function defined as
\begin{equation}
    p^k_{n_j}(t) = 
    \left\{ \begin{array}{ll}
        c^a_{n_j}, & \text{if} \; s^k_{n_j}(t) - s^k_{n_j}(t-1) = 1 \\
        c^e_{n_j}, & \text{if} \; s^k_{n_j}(t) - s^k_{n_j}(t-1) = -1 \\
       0, & \text{otherwise}
   \end{array} \right.
\label{eq:pen}
\end{equation}
{\bf Forwarding:} Let $\mc{L}^k$ be the set of links which are allowed to transmit VIPs of object k, determined by a routing policy. For each data object $\kin$ and each link $\abin^k$, choose
    \begin{equation}
        \mu^k_{ab}(t) =
        \left\{ \begin{array}{ll}
            C_{ba}, & \text{if} \; k = k^*_{ab}(t) \; \text{and} \; W^k_{ab}(t) > 0 \\
            0, & \text{otherwise}
        \end{array} \right.
    \label{eq:mul_flowchoice}
    \end{equation}
where,
    \begin{equation}
    \begin{split}
        W^k_{ab}(t) & \triangleq V^k_a(t) - V^k_b(t), \\
        k^*_{ab}(t) & \triangleq \argmax\limits_{\{k: \abin^k\}} W^k_{ab}(t),
    \end{split}
    \label{eq:mul_fwddefs}
    \end{equation}
{\bf Queue Evolution:} Update VIP counts according to
\begin{equation}
\begin{split}
    V^k_n(t+1) \leq & \Bigg(\Big(V^k_n(t) - \sum\limits_{\bin}\mu^k_{nb}(t)\Big)^+ + A^k_n(t) \\ 
    & + \sum\limits_{\ain}\mu^k_{an}(t) - \sum\limits_{\jin} r_{n_j} s^k_{n_j}(t) \Bigg)^+    
\end{split}
\label{eq:evo}
\end{equation}
where $(x)^+ \triangleq \max(x,0)$.
\label{alg:mul_pen}
\end{algorithm}
The caching subproblem of Algorithm~\ref{alg:mul_pen} defined by \eqref{eq:gap_obj}--\eqref{eq:gap_const3} ensures an optimal placement of objects across cache tiers with respect to both the value of objects and the costs of admissions and replacements.
The importance weight $\omega$ acts as a control parameter that lets us decide how much we value each side of this trade-off.
The penalty function \eqref{eq:pen} gets its name from the drift-plus-penalty approach and unifies the representation of cache utilization costs. Throughout the rest of the paper, we use the term {\it penalty} in reference to this function, in order to distinguish it from constants $c^a_{n_j}$ and $c^e_{n_j}$.
The last term in \eqref{eq:evo} represents VIPs removed via caching and lets us model the impact of the rate at which caches can satisfy requests, which directly affects network capacity and performance.

Forwarding according to \eqref{eq:mul_flowchoice} and \eqref{eq:mul_fwddefs} applies the backpressure algorithm \cite{tassiulas1990stability} to the VIP queues and forwards VIPs using links that maximize the queue backlog differential $W^k_{ab}$, which balances the VIP backlog around the network in the virtual plane. This, in turn, translates to balancing user demand for objects, thus enabling the implementation of a data plane forwarding policy that avoids congestion building up in parts of the network.

\subsection{Trade-off Optimality Analysis}
\label{subsec:vipanalysis}

We now provide an optimality analysis for our algorithm in the virtual plane, which consists of two steps: (i) defining the stability region of the VIP queue network and (ii) showing the optimal trade-off between queue backlog reduction and total penalty.

\subsubsection{Stability Region}
\label{subsubsec:vipcapacity}

As stated previously, our analysis of VIP queues in the virtual plane is based on the Lyapunov drift technique, which allows us to show that our algorithm can stabilize all VIP queues without a-priori knowledge of arrival rates. However, to show this property of the algorithm, a {\it stability region} of the network with respect to arrival rates, for which there exists some feasible policy that can make all VIP queues rate stable, must first be defined \cite{neelybook}. We formally define this region with the following theorem.

\begin{theorem}[Stability Region] The VIP stability region $\Lambda$ of the network $\mc{G} = (\mc{N},\mc{L})$, is the set $\Lambda$ consisting of all VIP arrival rates $\boldsymbol{\lambda} = (\lambda^k_n)_{\kin,\nin}$ such that the following holds
\begin{equation}
\begin{split}
    \lambda^k_n \leq \sum\limits_{\bin} f^k_{nb} & - \sum\limits_{\ain} f^k_{an} + \sum\limits_{\jin} r_{n_j} \betasumind, \\
    & \forall \nin, \; \kin, \; n \not = \mc{S}(k)
\label{eq:flowcaching}
\end{split}
\end{equation}
\begin{equation}
    0 \leq \beta_{n,i} \leq 1, \quad i = 1,\cdots,\sigma, \quad \nin
\label{eq:prob}
\end{equation}
\begin{equation}
    \betasum = 1, \quad \forall \nin
\label{eq:probsum}
\end{equation}
where $f^k_{ab} = \sum^{t}_{\tau=1} F^k_{ab}(\tau)/t$ denotes the time average VIP flow for object $k$ over link $(a,b)$ and $F^k_{ab}(t)$ is the number of VIPs of $k$ transmitted over $(a,b)$ during slot $t$ and satisfying the following for all $t \geq 1$.
\begin{equation}
    F^k_{ab}(t) \geq 0, F^k_{nn}(t) = 0, F^k_{\mc{S}(k)n}(t) = 0, \; \forall a,b,n \in \mc{N}, \kin,
\end{equation}
\begin{equation}
    F^k_{ab}(t) = 0, \; \forall a,b \in \mc{N}, \kin, (a,b) \not \in \mc{L}^k
\end{equation}
\begin{equation}
    \sum_{\kin} F^k_{ab}(t) \leq C_{ba}, \; \forall \abin
\end{equation}
$\mathcal{B}_{n,i}$ denotes the $i$-th one among $\sigma$ total number of possible cache placement sets at node $n$, and $\beta_{n,i}$ represents the fraction of time that objects are placed at $n$ according to $\mc{B}_{n,i}$. The elements of set $\mc{B}_{n,i}$ are pairs in the form of $(k,j)$, such that if $(k,j) \in \mc{B}_{n,i}$, object $k$ is cached in tier $j$.
\begin{proof}
Please see Appendix \ref{app:stability}.
\end{proof}
\label{thm:stability}
\end{theorem}

\subsubsection{Backlog-Penalty Bound Tradeoff}
\label{subsubsec:tradeoff}

We now show that our algorithm can achieve the optimal trade-off between queue backlog growth and total penalty, using the drift-plus-penalty approach.
We assume that request arrival processes are mutually independent for all nodes and objects, and are i.i.d. with respect to time slots. We also assume that there is a finite value $A^k_{n,max}$ such that $A^k_n(t) \leq A^k_{n,max}$ for all $n$, $k$ and $t$.

\begin{theorem}[Throughput and Penalty Bounds] Given the VIP arrival rate vector $\boldsymbol{\lambda} = (\lambda^k_n)_{\kin,\nin} \in int(\Lambda)$, if there exists $\boldsymbol{\epsilon} = (\epsilon^k_n)_{\nin, \kin} \succ \mathbf{0}$ such that $\boldsymbol{\lambda} + \boldsymbol{\epsilon} \in \Lambda$, then the network of VIP queues under Algorithm \ref{alg:mul_pen} satisfies
\begin{equation}
    \lim\limits_{T \rightarrow \infty} \frac{1}{T} \sum\limits^{T-1}_{t=0} \sum\limits_{\nin, \kin} \mathbbm{E}[V^k_n(t)] \leq \frac{NB}{\epsilon} + \frac{\omega}{2\epsilon} \minpen
\label{eq:vipbound}
\end{equation}
\begin{equation}
    \lim\limits_{T \rightarrow \infty} \frac{1}{T}\sum\limits^{T-1}_{t=0} \mathbbm{E}[p(t)] \leq \frac{2NB}{\omega} + \minpen
\label{eq:penbound}
\end{equation}
where $\epsilon \triangleq \textstyle \min_{\nin,\kin} \epsilon^k_n$, $p(t) = \sum_{\kin, \nin, \jin} p^k_{n_j}(t)$ is the total penalty accumulated across the network at slot $t$, $\Psi(\boldsymbol{\lambda})$ is the minimum long term average total penalty achievable by any feasible policy that makes all queues rate stable \cite{neelybook} and,
\begin{align}
    B \triangleq \frac{1}{2N} \sum_{\nin} & \bigg(\big(\textstyle \sum_{\bin}C_{nb}\big)^2 + 2\big(\textstyle \sum_{\bin}C_{nb}\big)\abs{\mc{K}}r_{n_1} \\ 
    & + \big(\textstyle \sum_{\kin}A^k_{n,max} + \textstyle \sum_{\ain}C_{an} + \abs{\mc{K}}r_{n_1}\big)^2 \bigg)
\end{align}
\begin{proof}
Please see Appendix \ref{app:dpp}.
\end{proof}
\label{thm:dpp}
\end{theorem}

\begin{figure}[!t]
  \begin{subfigure}[b]{0.22\textwidth}
    \includegraphics[width=\textwidth]{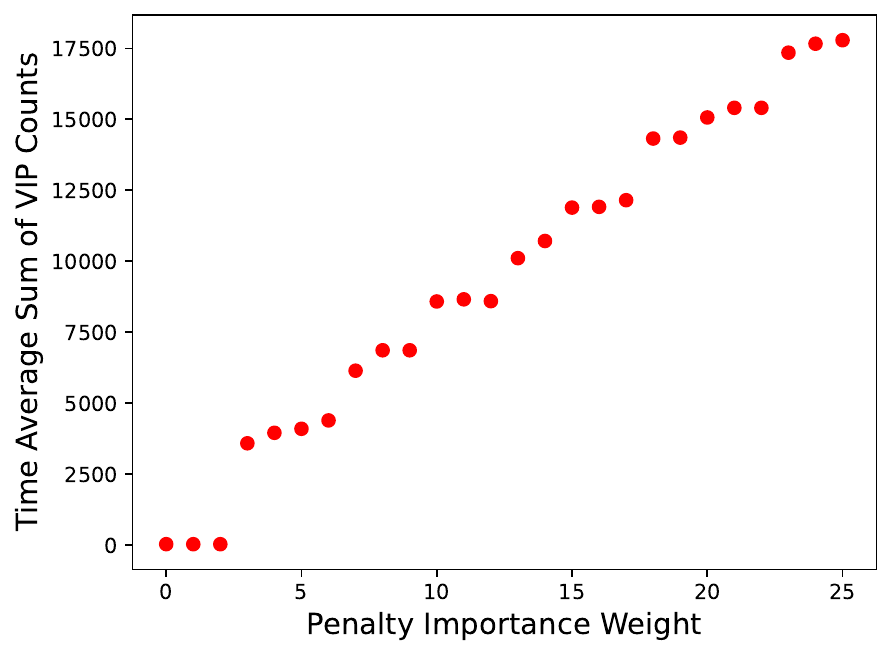}
    \caption{Sum of VIP counts}
    \label{fig:virtual_vip_avg}
  \end{subfigure}
  \begin{subfigure}[b]{0.215\textwidth}
    \includegraphics[width=\textwidth]{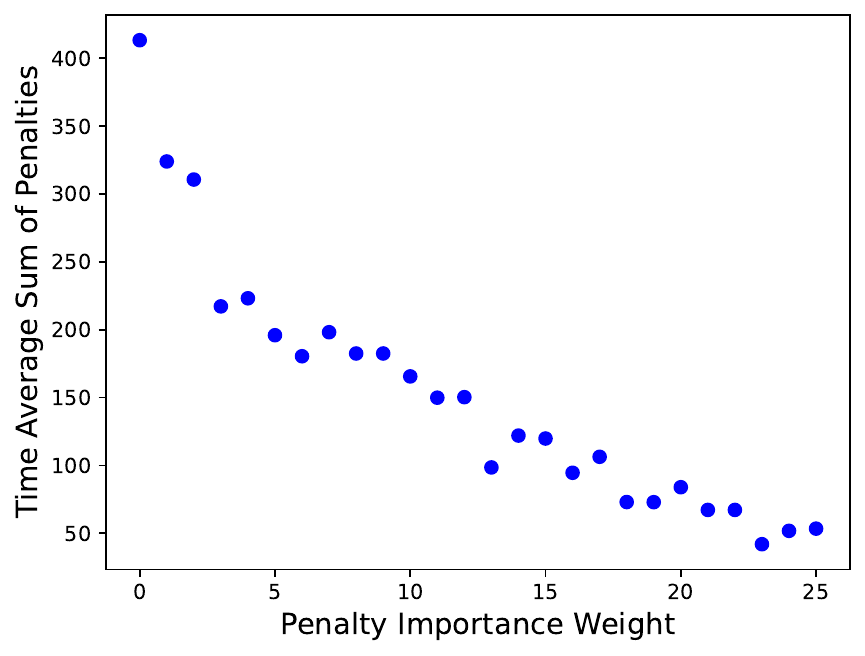}
    \caption{Sum of penalties}
    \label{fig:virtual_pen_avg}
  \end{subfigure}
  \caption{Long-term average of total VIP queue backlog and total penalty in the virtual plane (Abilene topology of Figure~\ref{fig:tops}, $\abs{\mc{N}}=1000$, $\sum_{\kin}\lambda^k_n=10$).}
  \label{fig:avgs}
\end{figure}

Theorem~\ref{thm:dpp} demonstrates the adjustable trade-off between queue backlogs and penalties: we can set $\omega$ arbitrarily large to push the time average penalty close to the minimum $\Psi(\boldsymbol{\lambda})$ at the cost of significantly increasing the average queue backlog and vice-versa. Figure~\ref{fig:avgs} visualizes this dynamic in the virtual plane.

\subsubsection{Optimal Solution to the Caching Subproblem}
\label{subsubsec:solution}

We have shown through Theorem~\ref{thm:dpp} that Algorithm~\ref{alg:mul_pen} optimizes the backlog-penalty trade-off. However, for Theorem~\ref{thm:dpp} to hold we must solve the caching subproblem defined by \eqref{eq:gap_obj}--\eqref{eq:gap_const3} exactly.
While this problem appears in the form of a generalized assignment problem (GAP), which is known to be NP-hard, we can show that it is in fact a rectangular assignment problem (RAP), by using a simple transformation of variables, which we present in the following lemma.

\begin{lemma}
Let $i$ be an integer in the set of integers $\mc{I}_n = \{1, 2, ..., \\  \sum_{\jin} L_{n_j}\}$. Let $b^k_{n_j}(t) = b^k_{n_i}(t)$ and $s^k_{n_j}(t) = s^k_{n_i}(t)$ if $i \in \mc{I}_{n_j} = \{1 + \sum^{j - 1}_{\ell = 1} L_{n_\ell}, ..., \sum^{j}_{\ell = 1} L_{n_\ell}\}$. Then, a solution to the following problem gives an equivalent solution to the problem defined by \eqref{eq:gap_obj}--\eqref{eq:gap_const3}.
\begin{align}
    \text{maximize}
        & \quad \sum\limits_{\kin} \sum\limits_{\iin} b^k_{n_i}(t) s^k_{n_i}(t) \label{eq:lsap_obj}\\
    \text{subject to} 
        & \quad \sum\limits_{\kin} s^k_{n_i}(t) \leq 1, \; \iin \label{eq:lsap_const1} \\
        & \quad \sum\limits_{\iin} s^k_{n_i}(t) \leq 1, \; \kin \label{eq:lsap_const2} \\
        & \quad s^k_{n_i}(t) \in \{0, 1\}, \; \kin, \; \iin \label{eq:lsap_const3}
\end{align}
\begin{proof}
    Since $|\mc{I}_{n_j}| = L_{n_j}$, $s^k_{n_j}(t) = 1$ can hold for at most $L_{n_j}$ objects. Therefore, \eqref{eq:lsap_const1} is equivalent to \eqref{eq:gap_const1}. Furthermore, since $s^k_{n_i} = 1$ can hold for at most one $i \in \mc{I}_n$, \eqref{eq:lsap_const2} is equivalent to \eqref{eq:gap_const2}.
\end{proof}
\label{lem:rap}
\end{lemma}

The problem defined by \eqref{eq:lsap_obj}--\eqref{eq:lsap_const3} has the form of a RAP and can be solved exactly in polynomial time using a modified Jonker-Volgenant algorithm \cite{volgenant1996linear,crouse2016implementing}. Since \eqref{eq:lsap_const1} is an inequality, in cases where $b^k_{n_i}(t) < 0$ for some $i \in \mc{I}_n$, we can obtain the optimal solution by treating \eqref{eq:lsap_const1} as an equality, then setting $s^k_{n_i}(t) = 0$ for all $i$ such that $b^k_{n_i}(t) < 0$.

It is important to point out that the equivalence of \eqref{eq:gap_obj}--\eqref{eq:gap_const3} to \eqref{eq:lsap_obj}--\eqref{eq:lsap_const3} stems from the assumption in our model that all objects have the same size. However, the problem formulated by \eqref{eq:gap_obj}--\eqref{eq:gap_const3} can be generalized to a model where objects have different sizes, in which case it would indeed be a GAP. Polynomial time methods that can approximate a solution to the GAP exist \cite{shmoys1993approximation,fleischer2006tight,feige2006approximation}. While the results of Theorem~\ref{thm:dpp} would not hold for an approximate solution, it may be possible to show that they would hold for arrival rates in the interior of a subset of $\Lambda$, the bounds of which depend on the factor of approximation.

%% file: Body/S5_data.tex
In this section, we develop practical policies that operate in the data plane and make forwarding and caching decisions by using Algorithm~\ref{alg:mul_pen} as a guideline.

{\it Caching.}
Implementing a caching policy in the data plane based on Algorithm~\ref{alg:mul_pen} requires certain adjustments due to a gap between the assumptions of the virtual plane and the constraints of the data plane.
Because we assumed that nodes have immediate access to any data object they decide to cache in the virtual plane, the virtual plane cache state variables $s^k_{n_j}(t)$ may not reflect the actual placement of objects in the data plane at any given time.
Furthermore, since VIPs for an object begin to be drained via caching immediately after the cache state for that object is set to 1, Algorithm~\ref{alg:mul_pen} leads to frequent shifts in cache distribution, as was also mentioned in \cite{VIPoriginal}. While the drift-plus-penalty approach allows us to mitigate this oscillatory behavior, it faces an issue in balancing the constant cost parameters that contribute to penalties against the large swings in VIP counts that can happen within a time slot. Having highlighted this gap, we present a practical caching policy in the data plane that is based on the VIP flows established by Algorithm~\ref{alg:mul_pen}.

We adopt the {\it cache score} metric devised for the stable caching algorithm presented in \cite{VIPoriginal}. Let $v^k_{ab}(t) \leq \mu^k_{ab}(t)$ be the actual number of VIPs for object $k$ transmitted over link $(a,b)$ during slot $t$ and let $T$ be the size of a sliding window. The cache score for object $k$ at node $n$ at time $t$ is defined as the average number of VIPs for object $k$ received by node $n$ over a sliding window of $T$ time slots prior to time slot $t$, expressed as follows.
\begin{equation}
    CS^k_n(t) = \frac{1}{T} \sum^t_{\tau = t - T + 1} \sum_{(a,n) \in \mc{L}^k} v^k_{an}(\tau)
    \label{eq:cache_scores}
\end{equation}
We then define the {\it cache benefit} metric, which balances the cache score against utilization costs, as follows.
\begin{equation}
    CB^k_{n_j}(t) = \begin{cases}
        \begin{array}{l}
             r_{n_j}(CS^k_n(t) - CS^{k'}_n(t)) - \omega(c^a_{n_j} + c^e_{n_j}), \text{if} \, j \, \text{is full} \\
             r_{n_j} CS^k_n(t) - \omega c^a_{n_j}, \; \text{otherwise}
        \end{array}
    \end{cases} 
    \label{eq:cache_benefit}
\end{equation}
where $k'_{n_j} = \argmin_{\{k \in \mc{K}_{n_j}\}} CS^k_n(t)$ with $\mc{K}_{n_j}$ denoting the set of currently cached objects across all tiers at node $n$ at a given time.

When a data object $k \not \in \mc{K}_{n_j}$ arrives at node $n$ during the time interval $[t,t)$, the caching policy first determines the cache tier which offers the highest cache benefit, i.e. $j^* = \argmax_{\{ \jin \}} CB^k_{n_j}(t)$. If $CB^k_{n_{j^*}}(t) > 0$, the object is admitted to tier $j^*$. If tier $j^*$ is full, object $k$ replaces object $k'_{n_j}$ defined above. If $j^*$ has free capacity, no replacements occur. We repeat this process as long as there are replacements, treating any replaced object as if it were a new data object arrival to see if there is benefit to migrating it to a different cache tier.
\begin{figure}[!t]
  \begin{subfigure}[b]{0.24\textwidth}
    \includegraphics[width=\textwidth]{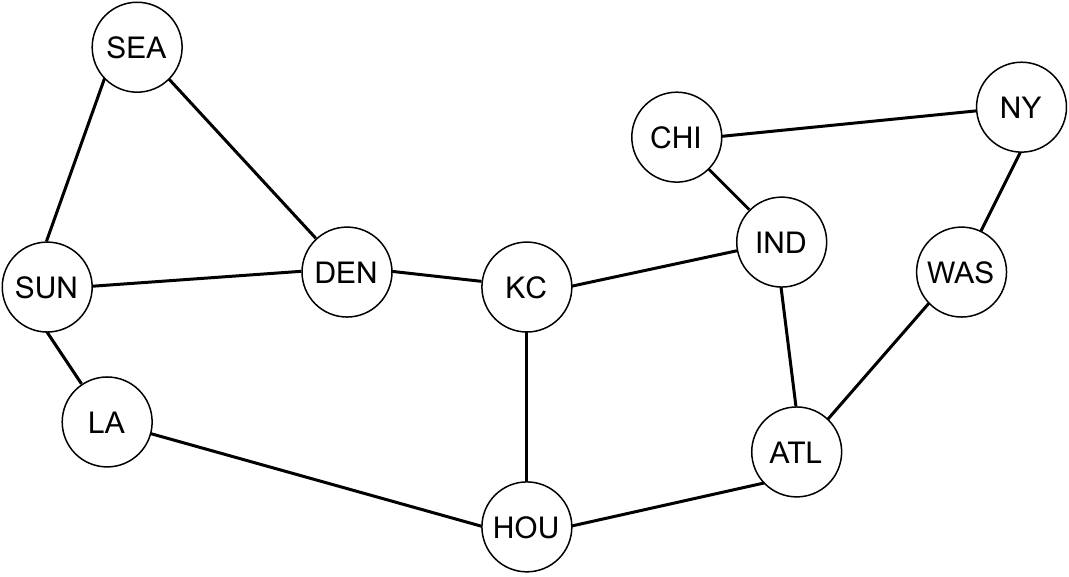}
    \caption{Abilene Network}
    \label{fig:tops_ab}
  \end{subfigure}
  \hfill
  \begin{subfigure}[b]{0.20\textwidth}
    \includegraphics[width=\textwidth]{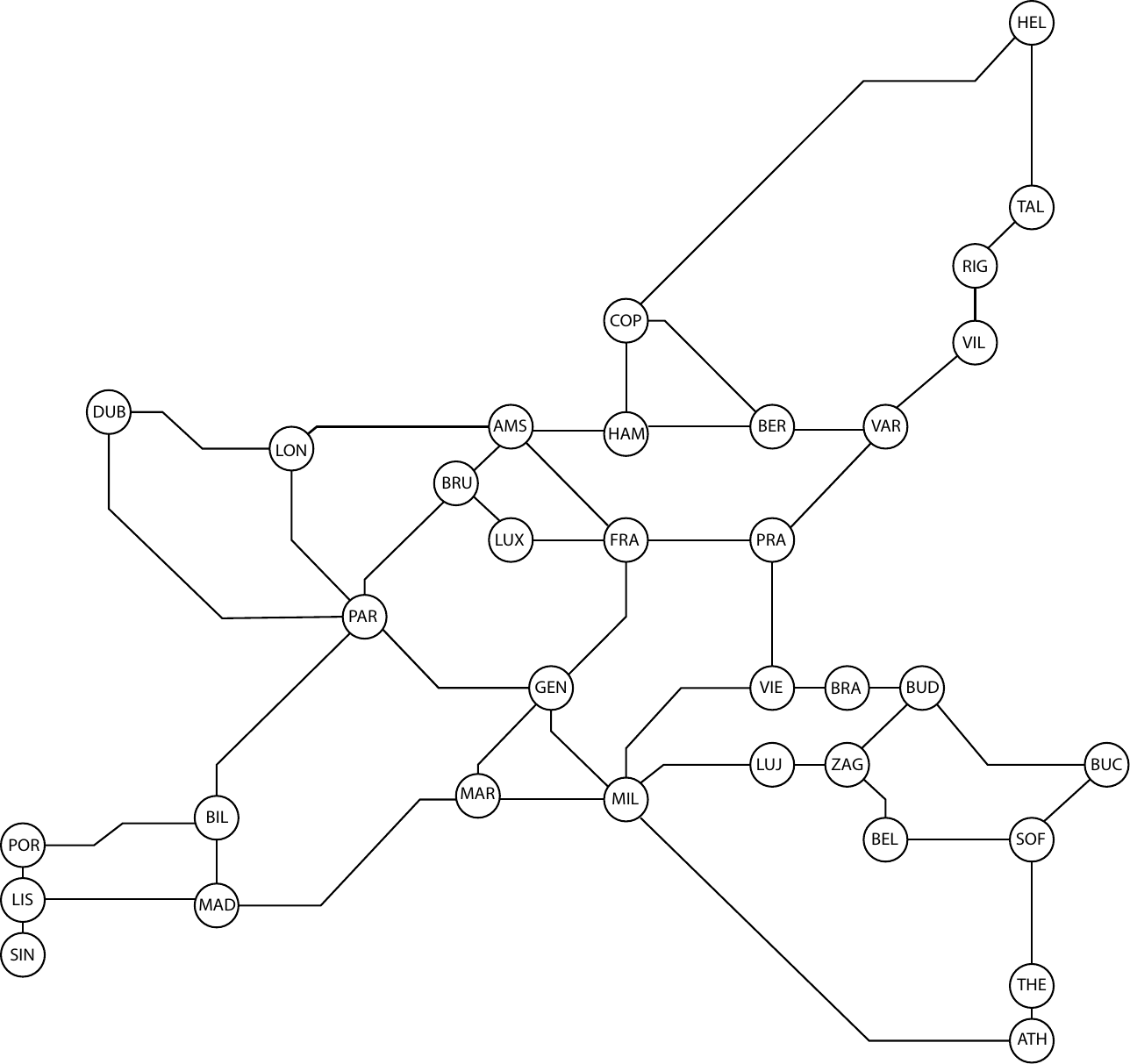}
    \caption{GEANT Network }
    \label{fig:tops_geant}
  \end{subfigure}
  \caption{Network Topologies}
  \label{fig:tops}
\end{figure}

{\it Forwarding.}
Implementing a forwarding policy in the data plane requires a similar, albeit simpler, adjustment to that of caching.
Although VIPs in the virtual plane are generated along with requests that arrive in the data plane, they are forwarded in batches periodically at each time slot, whereas each request in the data plane is forwarded as soon as it arrives. Therefore, we need a forwarding policy in the data plane that follows the flow patterns of VIPs in the virtual plane \cite{VIPoriginal}.
A request for object $k \not \in K_{n_j}$ that arrives at node $n$ is forwarded to node
\begin{equation}
    b^k_n(t) = \argmax_{\{ b:(n,b) \in \mc{L}^k \}} \frac{1}{T} \sum\limits^t_{t'=t-T+1} v^k_{nb}(t)
    \label{eq:dataplanefwd}
\end{equation}
Here, the sliding window is used once again, this time to represent the time-average behavior of the periodic VIP forwarding process in the virtual plane. This policy forwards requests on the most profitable links, as established by the flow patterns of VIPs in the virtual plane.

Note that the policies we describe above are expressed only in object level terms. In an actual implementation, data objects can consist of several {\it chunks}. Our policies would employ the following principles at the chunk level. If a data object is admitted to a cache tier, all its chunks must be admitted to the same tier. If a data object is evicted from a cache tier, all its chunks must be evicted from that tier. The forwarding decision described in \eqref{eq:dataplanefwd} is made when a request for the first chunk of object $k$ arrives at node $n$. Requests for subsequent chunks of $k$ are forwarded to the same node.

%% file: Body/S6_numerical.tex
\begin{figure*}[!t]
  \begin{subfigure}[b]{0.49\textwidth}
    \includegraphics[width=\textwidth]{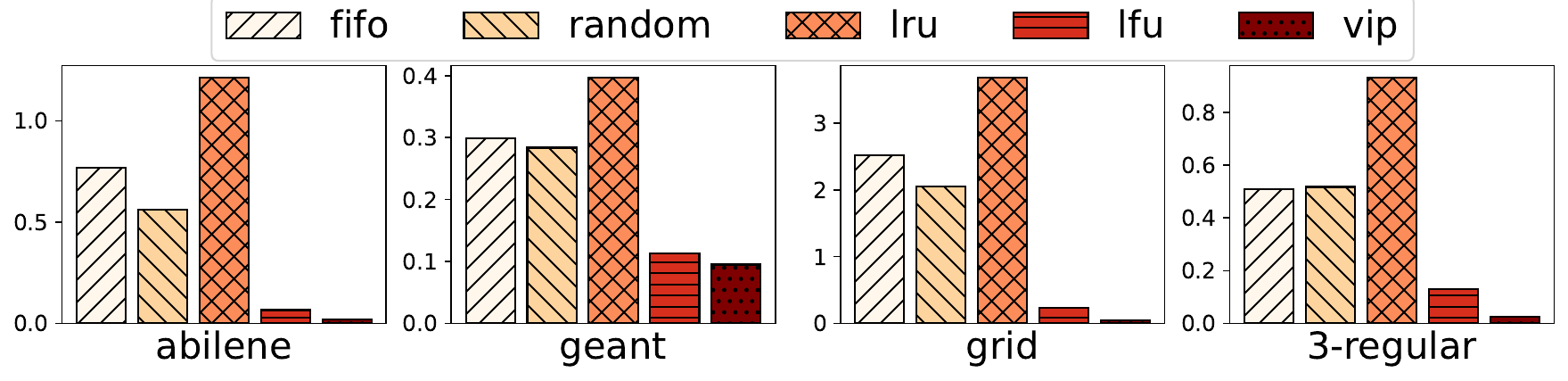}
    \caption{Total delay, as fraction of total delay without any caching}
    \label{fig:toptest_delay}
  \end{subfigure}
  \begin{subfigure}[b]{0.49\textwidth}
    \includegraphics[width=\textwidth]{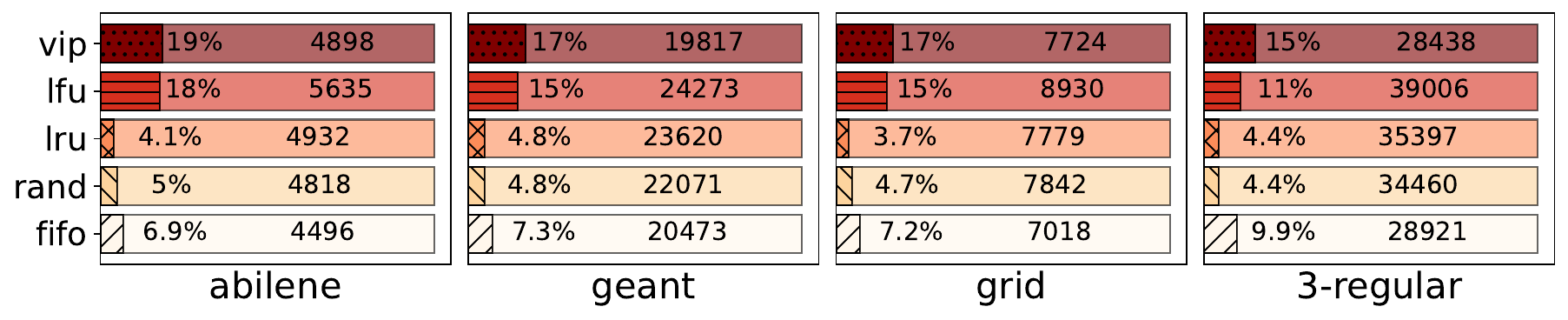}
    \caption{Cache hits, percentage of hits in the first tier on the left, total number of hits on the right}
    \label{fig:toptest_hits}
  \end{subfigure}
  \caption{Total delay and cache hits obtained by various policies across different topologies ($\lambda=10$, $L_{n_2}=100$, $\omega=0$).}
  \label{fig:toptest}
\end{figure*}
We now describe our experiment setting and results obtained for two test scenarios via simulations\footnote{The code for our simulations is available at \url{https://github.com/fvmutlu/multi-tier}.} over this setting.

\subsection{Experiment Setting}
We use the following topologies in our simulation experiments: the Abilene and Geant GN4-3N \cite{geant} network graphs shown in Figure~\ref{fig:tops}, a two dimensional $4 \times 4$ grid, and a 3-regular graph sampled uniformly. For all topologies, we assume that all nodes can be content sources, have caches and generate requests. For each data object $k$, a node is selected as the content source $\mc{S}(k)$ uniformly at random and independent of all other objects. We assume that node $\mc{S}(k)$ can produce enough copies of $k$ at any given time to satisfy all requests it receives for it. We apply a simple routing scheme on all topologies that allows each node $n$ to forward requests for object $k$ to any neighboring node that is its next hop on a shortest path (in number of hops) between $n$ and $\mc{S}(k)$.
In all topologies, each link has a capacity to transmit 10 data objects per second. For all settings, we consider a total of 1000 data objects. At each requester, requests arrive according to a Poisson process with a rate of $\lambda=10$. The object requested by each arrival is determined independently of previous arrivals according to a Zipf distribution with parameter 0.75. For all experiments, requests are generated for the first 100 seconds of the simulation, and a run completes once all requests are fulfilled. We use the same two cache tiers at each node: a fast but small tier with high utilization costs and a slower but larger tier with lower costs. The fast tier can store 5 objects and produce 20 objects per second. It has an admission cost of 4 and an eviction cost of 2. The slow tier can produce 10 objects per second, has an admission cost of 2 and an eviction cost of 1. Its capacity is left as a variable parameter for our experiments.
In addition to these parameters, we also define a write rate for each tier equal to its read rate. This parameter is not modeled in our approach, as our design aims to extract many cache hits for each cache admission and avoid frequent low-benefit replacements, making the write rate a negligible factor. We include this parameter in the simulations to incorporate the impact it could have in a real deployment. 
Note that the choices of numerical values stated here are not meant to replicate real network and cache device specifications, but rather to establish a simple object level setting that can capture the use case for our model and policies.

We run our virtual plane algorithm with a time slot length of 1 second, and use a window size $T$ of 100 slots to compute \eqref{eq:cache_scores} and \eqref{eq:dataplanefwd} for the data plane policies.
To establish baselines, we use four cache replacement policies: First In First Out (FIFO), Least Recently Used (LRU), Uniform Random (RAND) and Least Frequently Used (LFU). The first three replacement policies are adapted to the multi-tier model in "naive" ways, i.e. they do not account for performance and cost parameters of caches, and are implicitly paired with the Leave Copy Everywhere (LCE) admission policy, i.e. they admit all new (not already cached) data object arrivals:
\begin{itemize}[leftmargin=5mm]
    \item LRU: An arriving data object is admitted to the first cache tier and the LRU object in the tier is evicted if the tier is full. When an object is evicted from the first tier, it can be migrated to the second tier if there is available capacity there, or the LRU object in the second tier was requested less recently. An object evicted from the second tier cannot be migrated to the first tier and is evicted entirely from the node.
    \item FIFO: Both caches are operated as a single FIFO queue. When the object at the front of the first tier is evicted, it is migrated to the second tier. If the second tier is also full this causes the object at the front of the second tier to be evicted entirely from the node.
    \item RAND: A tier is selected to admit the arriving data object uniformly at random. If the selected tier is full, a random object from that tier is evicted entirely from the node.
\end{itemize}

We adapt LFU using the method in the practical caching policy we developed in Section~\ref{sec:s5_data}. To do so, in place of the cache score \eqref{eq:cache_scores}, we use the frequency measurement of LFU when computing the cache benefit \eqref{eq:cache_benefit}. We use an "ideal" LFU implementation which keeps a counter that tracks the number of accesses for all objects in the network. This multi-tier and cost-aware adaptation of LFU operates similarly to our caching policy in the data plane. As a result, when observing the performance differences between the two policies, we will be able to more directly evaluate the impact of having the virtual plane optimal algorithm drive decisions in the data plane.

We pair all four of the baseline caching policies described above with a Least Response Time (LRT) forwarding scheme, in which nodes keep track of the round trip delay of the last request sent over each outgoing link and choose the link with the smaller delay when more than one link is available to forward a request. For our data plane policy, we use this round trip delay measurement to break ties when \eqref{eq:dataplanefwd} yields more than one valid neighbor node.

\begin{figure*}[!t]
  \begin{subfigure}[b]{0.236\textwidth}
    \includegraphics[width=\textwidth]{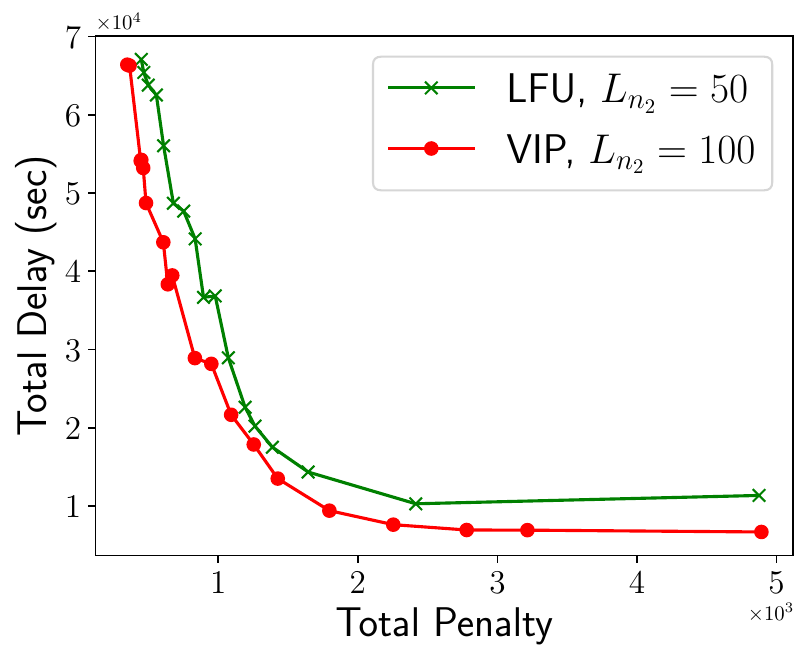}
    \caption{Abilene}
    \label{fig:pentest_abilene}
  \end{subfigure}
  \begin{subfigure}[b]{0.244\textwidth}
    \includegraphics[width=\textwidth]{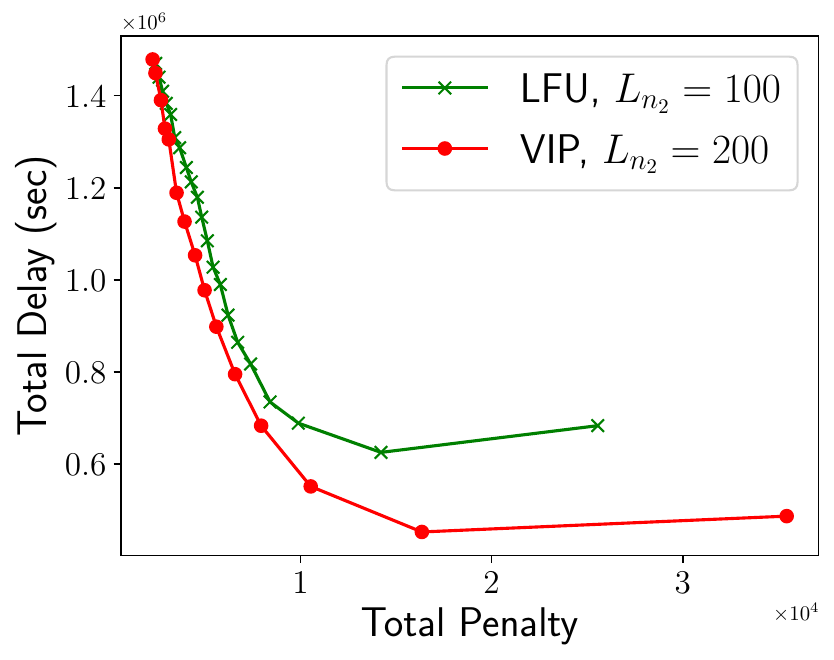}
    \caption{Geant}
    \label{fig:pentest_geant}
  \end{subfigure}
  \begin{subfigure}[b]{0.244\textwidth}
    \includegraphics[width=\textwidth]{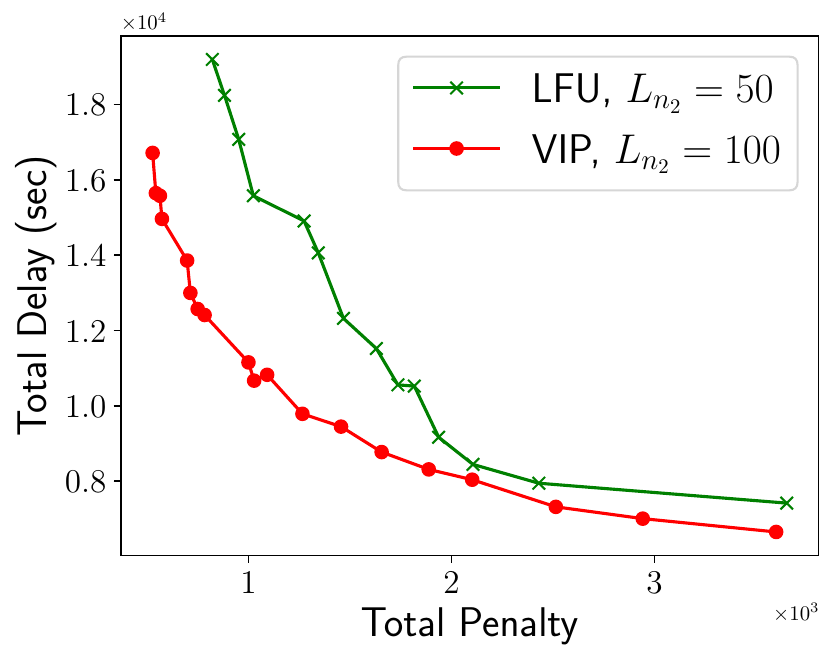}
    \caption{Grid}
    \label{fig:pentest_grid}
  \end{subfigure}
  \begin{subfigure}[b]{0.236\textwidth}
    \includegraphics[width=\textwidth]{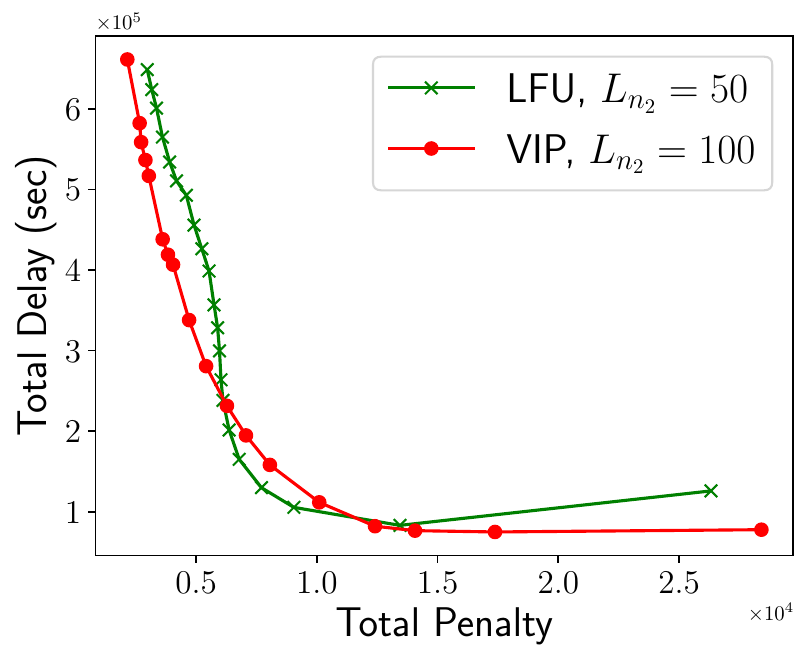}
    \caption{3-Regular}
    \label{fig:pentest_regular}
  \end{subfigure}
  \caption{Delay vs. penalty (tier 2 capacities for each policy listed in the legends)}
  \label{fig:pentest}
\end{figure*}

\subsection{Results}
We start with a simple scenario where penalties are disregarded by setting $\omega=0$ and the capacity of the second tier is fixed at 100 (denoted as $L_{n_2} = 100$). We present the results of this fixed-parameter scenario in Figure~\ref{fig:toptest}.
Figure~\ref{fig:toptest_delay} shows the total delay (the sum of delays experienced by all requests generated throughout the simulation run) achieved by the policies we test, as a fraction of the delay that would occur without any caching. It can be immediately seen that naive policies are performing poorly, even increasing the delay beyond that would be experienced without caching in certain cases. When we observe Figure~\ref{fig:toptest_hits}, which shows both the total number of cache hits and the distribution of these hits between the two tiers, we get a clearer picture. For these policies, the issue arises not from a lack of cache hits, but instead from an inability to balance the amount of requests served from each of the tiers. Furthermore, they perform many low-benefit replacements which exacerbates their poor performance. We can see this effect by observing the random strategy, which performs at most one replacement per data object arrival: while it has similar cache hit metrics with LRU, it achieves less delay in all topologies. Essentially, such naive policies trade queueing delays at network links with queueing delays at cache device controllers. In contrast, we can see that the adapted policies achieve much smaller delay with comparable amounts of total cache hits. In the grid topology, which exhibits the most drastic difference, the naive LRU policy increases total delay by 267\% (in comparison to no caching), while our VIP-based policy reduces delay by 95\%, which is a significant improvement even in comparison to the adapted LFU policy, which reduces delay by 76\%.

We now exclude the naive approaches and instead focus on the performance of our policy and the adapted LFU which, as we projected earlier and as demonstrated in Figure~\ref{fig:toptest}, has been performing closely to our policy. We explore the range of trade-offs for both policies by controlling $\omega$. We illustrate our results in Figure~\ref{fig:pentest} by plotting total delay against total penalty (the sum of penalties incurred by all caching decisions throughout the simulation run) to directly demonstrate the performance-cost trade-off. Note that different tier 2 capacities are used for the two policies, as we've experimented with different values of this parameter and picked the value yielding the best curve for each policy.

We can observe that our VIP-based policy outperforms LFU in all topologies except for a small region in the 3-regular topology.
While the adapted LFU policy behaves similarly to the VIP-based policy in the data plane, this performance difference between the two arises from the joint caching and forwarding optimization approach in the virtual plane that drives the VIP-based policy, which can take advantage of a larger tier 2 cache to obtain a better trade-off.

%% file: Body/S8_conclusion.tex
Leveraging recent advances in memory and storage technologies to scale cache capacities is becoming a necessity for high-performance data delivery networks. In this paper, we presented an object-level multi-tiered caching policy that can address the challenges in realizing this transition. We built our policy using an existing framework that provides powerful design tools for this challenge. We showed that our approach yields an optimal algorithm for the performance-cost trade-off in the virtual plane of this framework. We demonstrated through simulation experiments that the data plane policy driven by our algorithm in the virtual plane outperforms both traditional policies and an ideal LFU implementation that is adapted using the same principles as our data plane policy.

We conclude our work by underlining two potential future extensions.
Firstly, given the data delivery mechanics of our data plane model, our policy can run into issues where an object that has been previously cached in a slower tier sees significant increase in demand, but cannot be moved to a faster tier since requests for it are not forwarded upstream. To work around this, we could devise {\it proactive migration} schemes for the data plane, where objects can be moved between cache tiers independently of request arrival and data delivery processes. An example of a practical approach is to periodically poll the placement of objects across all tiers and carry out the most beneficial migrations that can be completed in a reasonable time frame.
Secondly, the assumption in the virtual plane that the readout rate of a cache tier is independent for each object does not translate well to the data plane since cache devices have a shared bandwidth. A way to amend this gap is to express the readout rates as dynamic parameters which are tied together by a total rate constraint. While the optimal selection of these dynamic parameters is an added problem dimension, it can be approached in the virtual plane using the backpressure-based tools provided by the framework.

%% file: Appendices/A1_stabilityregion.tex
Define the following for $\nin$, $\jin$ and $i=1,\cdots,\sigma$.\begin{equation*}
\begin{split}
    \mathcal{T}_{n,j,i} = \Bigg\{\tau \in \{1,\cdots,\tilde{t}\}: & s^k_{n_j}(\tau) = 1 \quad \forall \kjin \\
    & s^k_{n_j}(\tau) = 0 \quad \forall \kjout \Bigg\}
\end{split}
\end{equation*}
\begin{equation*}
\begin{split}
    \mathcal{T}_{n,i} = \Bigg\{\tau \in \{1,\cdots,\tilde{t}\}: & \sum\limits_{\jin} s^k_{n_j}(\tau) = 1 \quad \forall k: \exists \kjin \\
    & \sum\limits_{\jin} s^k_{n_j}(\tau) = 0 \quad \forall k: \forall \kjout \Bigg\}
\end{split}
\end{equation*}
Note that, due to the cache exclusivity assumption, $\sum_{\jin} \abs{\mathcal{T}_{n,j,i}} = \abs{\mathcal{T}_{n,i}}$. Then, defining $\beta_{n,i} = \abs{\mathcal{T}_{n,i}}/\tilde{t}$, we can prove \eqref{eq:prob} and \eqref{eq:probsum}.

Based on the evolution dynamics of VIP queues, we have that
\begin{equation}
    V^k_n(t) \geq \sum\limits^{t}_{\tau=1} A^k_n(\tau) + \sum\limits^{t}_{\tau=1} \sum\limits_{\ain} F^k_{an}(\tau) - \sum\limits^{t}_{\tau=1} \sum\limits_{\bin} F^k_{nb}(\tau) - \sum\limits_{\jin} r_{n_j} s^k_{n_j}(\tau)
\label{eq:a1_stability_2}
\end{equation}
Network stability, as defined in Lemma 1 of \cite{neely2005}, implies that there exists a finite $M$ such that $V^k_n(t) \leq M, \; \forall \nin,\kin$ holds infinitely often. Given an arbitrarily small value $\epsilon > 0$, there exists a slot $\Tilde{t}$ such that
\begin{equation}
    V^k_n(\Tilde{t}) \leq M, \quad \frac{M}{\Tilde{t}} \leq \epsilon, \quad \Bigg| \frac{\sum^{\Tilde{t}}_{\tau=1} A^k_n(\tau)}{\Tilde{t}} - \lambda^k_n \Bigg| \leq \epsilon
\label{eq:a1_stability_1}
\end{equation}
Then, by \eqref{eq:a1_stability_1} and \eqref{eq:a1_stability_2} we have
\begin{equation}
    \begin{split}
         \lambda^k_n - \epsilon \; \leq \; \frac{1}{\tilde{t}} \sum\limits ^{\tilde{t}}_{\tau=1} A^k_{n}(\tau)
         \; & \leq \; \frac{1}{\tilde{t}} V^k_{n}(\tilde{t}) + \frac{1}{\tilde{t}} \sum\limits ^{\tilde{t}}_{\tau=1} \sum\limits_{\bin} F^k_{nb}(\tau) \\
         & - \frac{1}{\tilde{t}}  \sum\limits ^{\tilde{t}}_{\tau=1} \sum\limits_{\ain} F^k_{an}(\tau) + \sum\limits_{\jin} r_{n_j} \frac{1}{\tilde{t}} \sum\limits ^{\tilde{t}}_{\tau=1} s^k_{n_j}(\tau)
    \end{split}
    \label{eq:a1_stability_3}
\end{equation}
Since $\sum\limits ^{\tilde{t}}_{\tau=1} s^k_{n_j}(\tau) = \sum\limits ^{\sigma}_{i=1} \abs{\mathcal{T}_{n,j,i}}\mathbf{1}_{[\kjin]}$, by \eqref{eq:a1_stability_3}, we have
\begin{equation}
    \lambda^k_n \; \leq \; \sum\limits_{\bin} f^k_{nb} - \sum\limits_{\ain} f^k_{an} + \sum\limits_{\jin} r_{n_j} \betasumind + 2\epsilon
\end{equation}
Thus, by letting $\epsilon \rightarrow 0$, we can prove \eqref{eq:flowcaching}, which defines the boundaries of the VIP network stability region $\Lambda$.

Now, we will further show that $\boldsymbol{\lambda} \in int(\Lambda)$ is sufficient for stability, i.e. there exists a feasible policy that can stabilize all queues given $\boldsymbol{\lambda} \in int(\Lambda)$. A feasible policy is one that, at each time slot $t$, yields a VIP forwarding allocation vector $(\mu^k_{ab}(t))_{\kin, \abin}$ and cache state vector $(s^k_{n_j}(t))_{\kin, \nin, \jin}$, which satisfy the following.
\begin{equation}
    \sum\limits_{\kin} \mu^k_{ab}(t) \leq C_{ab}, \; \text{for all} \; \abin
\label{eq:linkcapconst1}
\end{equation}
\begin{equation}
    \mu^k_{ab}(t) = 0, \; \text{for all} \; \about
\label{eq:linkcapconst2}
\end{equation}
\begin{equation}
    \sum\limits_{\kin} s^k_{n_j}(t) \leq L_{n_j}, \; \jin
\label{eq:cachecapconst1}
\end{equation}
\begin{equation}
    \sum\limits_{\jin} s^k_{n_j}(t) \leq 1, \; \kin
\label{eq:cachecapconst2}
\end{equation}
$\boldsymbol{\lambda} \in int(\Lambda)$ implies that there exists $\boldsymbol{\epsilon} = (\epsilon^k_n)_{\nin,\kin}$, where $\epsilon^k_n > 0$, such that $\boldsymbol{\lambda} + \boldsymbol{\epsilon} \in \Lambda$. Let $(f^k_{ab})_{\abin, \kin}$ and $\beta_{n,i}$ now denote the flow and cache placement set frequency variables associated with $\boldsymbol{\lambda} + \boldsymbol{\epsilon}$. Then, the following holds.
\begin{equation}
\begin{split}
    \lambda^k_n + \epsilon^k_n \leq \sum\limits_{\bin} f^k_{nb} - \sum\limits_{\ain} f^k_{an} + & \sum\limits_{\jin} r_{n_j} \betasumind, \\
    & \forall \nin, \kin, n \not = \mc{S}(k)
\end{split}
\label{eq:a1_stability_suff_1}
\end{equation}
We now construct a random forwarding policy. For every link $\abin$ such that $\sum_{\kin} f_{ab}^k > 0$, this policy transmits the VIPs of the single object $\Tilde{k}_{ab}$ which is chosen randomly to be $k$ with probability $f^k_{ab}/\sum_{\kin} f_{ab}^k$. Then, the number of VIPs that can be transmitted in slot $t$ is as follows:
\begin{equation}
    \Tilde{\mu}^k_{ab}(t) = 
    \begin{cases}
    \sum_{\kin} f_{ab}^k, & \text{if  } k = \Tilde{k}_{ab} \\
    0, & \text{otherwise}
    \end{cases}
\end{equation}
For every link $\abin$ such that $\sum_{\kin} f_{ab}^k = 0$, the policy sets $\Tilde{\mu}^k_{ab}(t) = 0$ for all $\kin$. Thus, we have
\begin{equation}
    \mathbbm{E}[\Tilde{\mu}^k_{ab}(t)] = f^k_{ab}
\label{eq:a1_stability_suff_2}
\end{equation}
Next, we construct a randomized caching policy. For every node $n$, this policy caches the distribution $\tilde{\mathcal{B}}_n$, where $\tilde{\mathcal{B}}_n$ is chosen randomly to be $\mathcal{B}_{n,i}$ with probability $\beta_{n,i}$. Then, the caching state of $k$ for tier $j$ during slot $t$ is as follows:
\begin{equation}
    \tilde{s}^k_{n_j}(t) = 
    \begin{cases}
    1, & \text{if  } (k,j) \in \tilde{\mathcal{B}}_n \\
    0, & \text{otherwise}
    \end{cases}
\label{eq:a1_stability_cachingstate}
\end{equation}
Thus, we have
\begin{align}
\begin{split}
    \mathbbm{E} \Big[\sum\limits_{\jin} r_{n_j} \tilde{s}^k_{n_j}(t)\Big] 
    & = \sum\limits_{\jin} r_{n_j}
    \mathbbm{E}[\tilde{s}^k_{n_j}(t)] \\
    & = \sum\limits_{\jin} r_{n_j} \betasumind
\label{eq:a1_stability_suff_3}
\end{split}
\end{align}
Then, by \eqref{eq:a1_stability_suff_1}, \eqref{eq:a1_stability_suff_2} and \eqref{eq:a1_stability_suff_3}, we have
\begin{equation}
\begin{split}
    \lambda^k_n + \epsilon^k_n & \leq \mathbbm{E} \Big[\Big(\sum\limits_{\bin} \tilde{\mu}^k_{nb}(t) - \sum\limits_{\ain} \tilde{\mu}^k_{an}(t) + \sum\limits_{\jin} r_{n_j} \tilde{s}^k_{n_j}(t) \Big)\Big] \\
    & = \sum\limits_{\bin} f^k_{nb} - \sum\limits_{\ain} f^k_{an} + \sum\limits_{\jin} r_{n_j} \betasumind 
\end{split}
\label{eq:a1_stability_loynes}
\end{equation}
Equation \eqref{eq:a1_stability_loynes} shows that the service rate is greater than the arrival rate for a randomized feasible policy given $\boldsymbol{\lambda} \in int(\Lambda)$, which proves that this condition is sufficient for stability. \qed

%% file: Appendices/A2_minimaldpp.tex
Define the Lyapunov function as

\begin{equation}
    \mc{L}(\mathbf{V}(t)) = \sum\limits_{\nin, \kin} (V^k_n(t))^2
\end{equation}

Then the Lyapunov drift at slot $t$ is given by

\begin{equation}
    \drift = \mathbbm{E}[\mathcal{L}(\mathbf{V}(t+1))-\mc{L}(\mathbf{V}(t))|\mathbf{V}(t)]
\label{eq:drift}
\end{equation}

Taking the square of both sides of \eqref{eq:evo}, we have the following

\begin{align*}
    & \big(V^k_n(t+1)\big)^2 \leq \big( V^k_n(t) \big)^2 + \bigg(\flownb\bigg)^2 - 2 V^k_n(t) \flownb \\
    & + 2 V^k_n(t) \bigg(A^k_n(t) + \flowan \bigg) \\
    & - 2 \bigg( V^k_n(t) - \flownb \bigg)^+ \bigg( \rosumtier \bigg) \\
    & + \bigg(A^k_n(t) + \flowan - \rosumtier \bigg)^2
\end{align*}

The terms can be collected as follows
\begin{align*}
    & \big(V^k_n(t+1)\big)^2 - (V^k_n(t))^2 \\ 
    & \leq \bigg(\flownb\bigg)^2 + 2\flownb \rosumtier
\end{align*}
\begin{align*}
    & + \bigg(A^k_n(t) + \flowan + \rosumtier \bigg)^2 + 2V^k_n(t)A^k_n(t) \\
    & - 2V^k_n(t)\bigg(\flownb - \flowan \bigg) - 2V^k_n(t) \rosumtier
\end{align*}
We can then take the sum over $\nin, \kin$ to obtain
\begin{align*}
    & \mc{L}(\mathbf{V}(t+1) - \mc{L}(\mathbf{V}(t)) \\
    & \leq \sum\limits_{\nin, \kin} \bigg(\flownb\bigg)^2 + 2 \sum\limits_{\nin, \kin} \flownb \rosumtier \\
    & + 2 \sum\limits_{\nin, \kin} \bigg(A^k_n(t) + \flowan + \rosumtier \bigg)^2 \\
    & + 2 \sum\limits_{\nin, \kin} V^k_n(t) A^k_n(t) - 2 \sum\limits_{\abin, \kin} \mu^k_{ab}(t) \bigg(V^k_a(t) - V^k_b(t) \bigg) \\
    & - 2 \sum\limits_{\nin, \kin} V^k_n(t) \rosumtier
\end{align*}
We now define the following constants: $\mu^{out}_{n,max} \triangleq \sum_{\bin}C_{nb}$, $\mu^{in}_{n,max} \triangleq \sum_{\ain}C_{an}$, $A_{n,max} \triangleq \sum_{\kin}A^k_{n,max}$, $r_{n,max} \triangleq Kr_{n_1}$.
Based on these definitions, the following hold.
\begin{align}
    & \sum\limits_{\kin} \bigg( \flownb \bigg)^2 \leq (\mu^{out}_{n,max})^2, \label{eq:maxdef1} \\
    & \begin{aligned}
    & \sum\limits_{\kin} \bigg( A^k_n(t) + \flowan + \rosumtier \bigg)^2 \\
    & \leq (A_{n,max} + \mu^{in}_{n,max} + r_{n,max})^2, \label{eq:maxdef2}
    \end{aligned} \\
    & \sum\limits_{\kin} \flownb \rosumtier \leq \mu^{out}_{n,max} r_{n,max} \label{eq:maxdef3}
\end{align}
Then, by \eqref{eq:maxdef1}-\eqref{eq:maxdef3} and the definition of $B$ we have
\begin{equation}
\begin{split}
    & \mc{L}(\mathbf{V}(t+1) - \mc{L}(\mathbf{V}(t)) \leq 2NB + 2 \sum\limits_{\nin, \kin} V^k_n(t) A^k_n(t) \\
    & - 2 \sum\limits_{\abin, \kin} \mu^k_{ab}(t) \bigg(V^k_a(t) - V^k_b(t) \bigg) \\
    & - 2 \sum\limits_{\nin, \kin} V^k_n(t) \rosumtier
\end{split}
\label{eq:drift_ineq3}
\end{equation}
Taking conditional expectations on both sides of \eqref{eq:drift_ineq3}, we have
\begin{equation}
\begin{split}
    & \drift \leq 2NB + 2\sum\limits_{\nin, \kin} V^k_n(t)\lambda^k_n(t) \\
    & - 2\mathbbm{E}\Bigg[\sum\limits_{\abin, \kin} \mu^k_{ab}(t) \bigg(V^k_a(t) - V^k_b(t) \bigg) \bigg| \mathbf{V}(t)\Bigg] \\
    & - 2\mathbbm{E}\Bigg[\sum\limits_{\nin, \kin} V^k_n(t)\rosumtier \bigg| \mathbf{V}(t)\Bigg]
\end{split}
\label{eq:drift_ineq_main}
\end{equation}

Now, let $\svec = (s^k_{n_j}(t))_{\kin, \nin, \jin}$ and $\hat{p}(\textbf{\textit{s}}(t)) = p(t)$, then add the penalty function to both sides of \eqref{eq:drift_ineq_main} to get the drift-plus-penalty inequality.

\begin{equation}
\begin{split}
    & \drift + \omega \pen \leq 2NB + 2\sum\limits_{\nin, \kin} V^k_n(t)\lambda^k_n(t) \\
    & - 2\mathbbm{E}\Bigg[\sum\limits_{\abin, \kin} \mu^k_{ab}(t) \bigg(V^k_a(t) - V^k_b(t) \bigg) \bigg| \mathbf{V}(t)\Bigg] \\
    & - 2\mathbbm{E}\Bigg[\sum\limits_{\nin, \kin} V^k_n(t)\rosumtier \bigg| \mathbf{V}(t)\Bigg] \\
    & + \omega \penhat
\end{split}
\end{equation}

Now, let $\tilde{\mu}^k_{ab}(t)$ and $\tilde{s}^k_{n_j}(t)$ refer to all feasible VIP forwarding allocation rates and cache state variables respectively. Then, given $\mu^k_{ab}(t)$ and $s^k_{n_j}(t)$ obtained via Algorithm \ref{alg:mul_pen}, the following holds.
\begin{equation}
\begin{split}
    & \drift + \omega \pen \leq 2NB + 2\sum\limits_{\nin, \kin} V^k_n(t)\lambda^k_n(t) \\
    & - 2\mathbbm{E}\Bigg[\sum\limits_{\abin, \kin} \mu^k_{ab}(t) \bigg(V^k_a(t) - V^k_b(t) \bigg) \bigg| \mathbf{V}(t)\Bigg] \\
    & - 2\mathbbm{E}\Bigg[\sum\limits_{\nin, \kin} V^k_n(t)\rosumtier \bigg| \mathbf{V}(t)\Bigg] \\
    & + \omega \penhat \\
    & \leq 2NB + 2\sum\limits_{\nin, \kin} V^k_n(t)\lambda^k_n(t) \\
    & - 2\mathbbm{E}\Bigg[\sum\limits_{\abin, \kin} \tilde{\mu}^k_{ab}(t) \bigg(V^k_a(t) - V^k_b(t) \bigg) \bigg| \mathbf{V}(t)\Bigg] \\
    & - 2\mathbbm{E}\Bigg[\sum\limits_{\nin, \kin} V^k_n(t)\rosumtiertilde \bigg| \mathbf{V}(t)\Bigg] \\
    & + \omega \pentilde
\end{split}
\end{equation}
This is because Algorithm 1 maximizes the following two quantities over all feasible $\tilde{s}^k_{n_j}(t)$ and $\tilde{\mu}^k_{ab}(t)$, by \eqref{eq:gap_obj}--\eqref{eq:gap_const3} and by (\ref{eq:mul_flowchoice})--(\ref{eq:mul_fwddefs}) respectively.
\begin{align*}
    & \sum\limits_{\nin, \kin} V^k_n(t)\bigg(\flownbtilde - \flowantilde \bigg) \\
     = & \sum\limits_{\abin, \kin} \tilde{\mu}^k_{ab}(t) \bigg(V^k_a(t) - V^k_b(t) \bigg),
\end{align*}
\begin{align*}
    \bigg(\sum\limits_{\nin, \kin} V^k_n(t)\rosumtiertilde\bigg) - \omega \hat{p}^k_{n_j}(\tilde{s}^k_{n_j}(t))
\end{align*}

Then, using \eqref{eq:a1_stability_loynes}, we have
\begin{align*}
    & \drift + \omega \pen \\
    & \leq 2NB + 2\sum\limits_{\nin, \kin} V^k_n(t)\lambda^k_n(t) - 2\sum\limits_{\nin, \kin} V^k_n(t) \\
    & \times \mb{E}\Bigg[ \bigg(\flownbtilde - \flowantilde + \rosumtiertilde \bigg) \bigg| \mathbf{V}(t) \Bigg] \\
    & + \omega \pentilde
\end{align*}
\begin{align*}
    & \leq 2NB + 2\sum\limits_{\nin, \kin} V^k_n(t)\lambda^k_n(t) \\
    & - 2\sum\limits_{\nin, \kin} V^k_n(t) (\lambda^k_n + \epsilon^k_n) + \omega \pentilde \\
    & \leq 2NB - 2\sum\limits_{\nin, \kin} \epsilon^k_n V^k_n(t) + \omega \pentilde \\
    & \leq 2NB - 2 \epsilon \sum\limits_{\nin, \kin} V^k_n(t) + \omega \pentilde
\end{align*}
Now, assume that, for a given arrival rate vector $\boldsymbol{\lambda}$, at time $t$, the policy that achieves $\minpen$ has the cache state variable vector $\svecstar$ and $\mathbbm{E}[\hat{p}(\svecstar)] = p^*(t)$. Therefore, we have
\begin{equation}
\begin{split}
    & \drift + \omega \pen \\
    & \leq 2NB - 2 \epsilon \sum\limits_{\nin, \kin} V^k_n(t) + \omega \penstar \\
    & = 2NB - 2 \epsilon \sum\limits_{\nin, \kin} V^k_n(t) + \omega p^*(t)
\end{split}
\end{equation}
Using (\ref{eq:drift}) and the law of iterated expectations we have
\begin{equation}
\begin{split}
    & \mathbbm{E}[\drift] + \omega \mathbbm{E}[\pen] \\
    & = \mathbbm{E}[\mathbbm{E}[\mathcal{L}(\mathbf{V}(t+1))-\mc{L}(\mathbf{V}(t))|\mathbf{V}(t)]] + \omega \mathbbm{E}[\pen] \\
    & = \mathbbm{E}[\mc{L}(\mathbf{V}(t+1))]-\mathbbm{E}[\mc{L}(\mathbf{V}(t))] + \omega \mathbbm{E}[p(t)] \\
    & \leq 2NB - 2 \epsilon \sum\limits_{\nin, \kin} \mathbbm{E}[V^k_n(t)] + \omega \minpen
\end{split}
\end{equation}
Summing both sides over $t \in \{1, 2, ..., T \}$ for some positive integer T, we obtain the following.
\begin{equation}
\begin{split}
    & \mathbbm{E}[\mc{L}(\mb{V}(T+1))] - \mathbbm{E}[\mc{L}(\mb{V}(1))] + \omega \sum\limits^{T}_{t=1} \mathbbm{E}[p(t)] \\
    & \leq 2NBT - 2\epsilon \sum\limits^{T}_{t=1} \sum\limits_{\nin, \kin} \mathbbm{E}[V^k_n(t)] + \omega T \minpen
\label{eq:Tsum}
\end{split}
\end{equation}
Note that the values of $V^k_n(t)$, $\mc{L}(\mb{V}(t))$ and $p(t)$ for any given $k$, $n$, and $t$ are all non-negative. Recall also that $\epsilon > 0$ and $\omega \geq 0$. Therefore, we can arrange the above into two inequalities:
\begin{equation}
\begin{split}
    \frac{1}{T}\sum\limits^{T}_{t=1} \mathbbm{E}[p(t)] & \leq \frac{2NB}{\omega} + \frac{\mathbbm{E}[\mc{L}(\mb{V}(1))]}{\omega T} + \minpen \\
    \frac{1}{T} \sum\limits^{T}_{t=1} \sum\limits_{\nin, \kin} \mathbbm{E}[V^k_n(t)] & \leq \frac{NB}{\epsilon} + \frac{\mathbbm{E}[\mc{L}(\mb{V}(1))]}{2 \epsilon T} + \frac{\omega}{2 \epsilon} \minpen
\end{split}
\end{equation}
If we take the limit as $T \rightarrow \infty$ for both inequalities, we reach the two final bounds.
\begin{equation}
    \lim\limits_{T \rightarrow \infty} \frac{1}{T}\sum\limits^{T}_{t=1} \mathbbm{E}[p(t)] \leq \frac{2NB}{\omega} + \minpen
\end{equation}
\begin{equation}
    \lim\limits_{T \rightarrow \infty} \frac{1}{T} \sum\limits^{T}_{t=1} \sum\limits_{\nin, \kin} \mathbbm{E}[V^k_n(t)] \leq \frac{NB}{\epsilon} + \frac{\omega}{2 \epsilon}\minpen
\end{equation}
\qed